\documentclass[fleqn,twoside,twocolumn,nofootinbib,showkeys]{revtex4} % Specifies the document class %,unsortedaddress
\usepackage[sec,nocpr]{ujp} % \usepackage[]{ujp} for cyrillic, \usepackage[web]{ujp} for web
\def\dilog{\mathop{\rm dilog}\nolimits}
\def\arcth{\mathop{\rm arcth}\nolimits}
\def\th{\mathop{\rm th}\nolimits}

\begin{document}
\title[Effective Mass of $^{4}$He Atom]%колонтитул
{EFFECTIVE MASS OF \boldmath$^{4}$He ATOM\\ IN SUPERFLUID AND NORMAL PHASES }%
\author{I.O. Vakarchuk}%1 автор
\affiliation{Ivan Franko National University of L'viv}%
\address{12, Dragomanov Str., Lviv 79005, Ukraine}
\email{chair@franko.lviv.ua}
\author{O.I.~Hryhorchak}%1 автор
\affiliation{Ivan Franko National University of L'viv}%
\address{12, Dragomanov Str., Lviv 79005, Ukraine}
%\email{HrOrest@gmail.com}
\author{V.S.~Pastukhov}%1 автор
\affiliation{Ivan Franko National University of L'viv}%
\address{12, Dragomanov Str., Lviv 79005, Ukraine}
%\email{@gmail.com}
\author{R.O.~Prytula}%1 автор
\affiliation{Ivan Franko National University of L'viv}%
\address{12, Dragomanov Str., Lviv 79005, Ukraine}
%\email{@gmail.com}

\udk{538.941}  \razd{\secvi}

\autorcol{I.O.\hspace*{0.7mm}Vakarchuk, V.S.\hspace*{0.7mm}Pastukhov
О.І.\hspace*{0.7mm}Hryhorchak, R.O.\hspace*{0.7mm}Prytula}

\setcounter{page}{29}%

\begin{abstract}
The formula for the temperature dependence of the effective mass of a $^{4}%
$He atom in the superfluid and normal phases is obtained.\,\,This
expression for the effective mass allows one to eliminate infra-red
divergences, being applicable at all temperatures, except for a
narrow fluctuation region 0.97~$\lesssim T/T_{\rm c}\leq1$.\,\,In
the high and low temperature limits, as well as in the
interactionless limit, the obtained expression reproduces the well
known results.\,\,The temperature dependence of the heat capacity
and the phase transition temperature $T_{\rm c}\approx$~2.18~K are
calculated, by using the formula obtained for the effective
mass.\,\,In the framework of the approach proposed in this work, the
small critical index $\eta$ is determined in the random phase
approximation.\,\,The obtained value corresponds to the well known
result.
\end{abstract}


\maketitle

\section{Introduction}

The idea that the transition of liquid $^{4}$He into the superfluid
state is a manifestation of the Bose--Einstein condensation was put
forward for the first time by F.~London \cite{London}.\,\,It was the
\textquotedblleft proximity\textquotedblright\ of the Bose
condensation temperature in an ideal gas with helium parameters to
the transition temperature in real $^{4}$He that suggested him this
idea.\,\,Although this interpretation of the phase transition is not
free from difficulties \cite{Pashitskii}, it correctly describes, in
general, modern experiments with cooled gases
\cite{DalfovoGiorginiPitaevskiiStringari,BlochDalibardZwerger}.

The problems in the theory of liquid $^{4}$He, which remain
unresolved till now, include the calculation of corresponding
thermodynamic functions in the whole temperature interval and the
calculation of the transition temperature into the superfluid state,
which would agree with the experimental value.\,\,At the qualitative
level, a reduction of the critical temperature was substantiated by
R.\,\,Feynman \cite{Feynman53}, who introduced the concept of
effective particle mass.\,\,For intuitive reasons, he came to a
conclusion that, owing to the interaction between particles, the
effective mass has to exceed the atomic one.\,\,This conclusion is
also valid for two-dimensional systems \cite{PieriStrinatiTifrea}.

However, there exists a competing mechanism.\,\,The repulsion at
short distances effectively increases the system density and,
consequently, should increase the Bose condensation
temperature.\,\,This conclusion is confirmed by the results of
theoretical calculations \cite{Stoof,
BaymBlaizotHolzmannLaloeVautherin, HolzmannKrauth,
ArnoldMooreTomasik, Kastening} and Monte-Carlo simulation carried
out for the model of weakly non-ideal Bose gas
\cite{KashurnikovProkofevSvistunov, ArnoldMoore}.\,\,In order to put
the experimental results obtained in the $^{4}$He-Vycor system in
correspondence with the results of theoretical calculations, the
effect of atomic mass renormalization and a shift associated with
the repulsive part of the interparticle interaction have to be taken
into account simultaneously
\cite{ReppyCrookerHebralCorwinHeZassenhaus}.

In the literature, the value of effective mass at low temperatures
was mainly analyzed \cite{Samulski, visnyk93,Vakar96,Vakar97,
Rovenchak03}.\,\,In works \cite{Lindenau99,Lindenau02}, the
corresponding temperature dependence was obtained within the
variational approach.\,\,The properties of helium in the normal
phase were studied in works \cite{Gernoth08, Gernoth09}, where the
effective mass of particles was used as a fitting parameter to put
the calculated structural functions in agreement with experimental
curves.

A method of calculation of the effective mass of $^{4}$He atoms in
the liquid phase was demonstrated in work \cite{VP2008}.\,\,It
allows infra-red divergences typical of the phase transition theory
to be eliminated.\,\,A shortcoming of the proposed approach is the
poorly substantiated extrapolation of the \textquotedblleft
seed\textquotedblright\ effective mass onto a wide temperature
interval (proceeding from the corresponding expression obtained for
the zero temperature) and an incorrect behavior of the obtained
effective mass in the critical region.

Another approach to the calculation of the effective mass was
proposed in work \cite{Past2012}.\,\,In its framework, the
temperature dependence of the heat capacity was determined.\,\,The
result obtained turned out in much better agreement with
experimental data than the results of calculations on the basis of a
\textquotedblleft bare\textquotedblright\ mass.\,\,However, the
expression for the effective mass, which was obtained in the
framework of this approach, does not exclude the mentioned infra-red
divergences, because the ideology  of the effective mass
calculations was not oriented to this purpose.

This work is aimed at finding such an expression for the effective
mass, which would eliminate infra-red divergences and reproduce a
correct behavior in a vicinity of the critical point (excluding,
maybe, a narrow fluctuation region).\,\,At the same time, it should
be better substantiated theoretically in a wide temperature
region.\,\,Another task consisted in obtaining the temperature
dependence of the heat capacity with the use of the new effective
mass and in comparing it with the previous results.

\section{General Formulas}

While calculating the heat capacity for a many-boson system, let us
use the expression for its internal energy in the pair correlation
approximation \cite{VP2008,Past2012}.\,\,The expression can be
obtained by averaging the Hamiltonian with the density matrix found
in work \cite{Vakarchuk04}:
 %1
 \[
E=N\frac{mc^2}{2}+\sum\limits_{{\bf q}\neq0}
\frac{\bar{\varepsilon}_q}{\bar{z}_0^{-1}e^{\beta\bar{\varepsilon}_q}-1}\,+
 \]\vspace*{-7mm}
 \[
+\,\frac{1}{2}\frac{\bar{m}}{m}\sum\limits_{{\bf
q}\neq0}\frac{\lambda_q}{1+\lambda_q\bar{S}_0(q)}
 \frac{\partial \bar{S}_0(q)}{\partial\beta}\,+
 \]%\vspace*{-7mm}
 \[
+\,\frac{1}{4}\sum\limits_{{\bf
q}\neq0}\varepsilon_q\left(\lambda_q^2+\alpha_q^2-1\right)S(q)\,+
 \]\vspace*{-7mm}
 \[
+\,\frac{1}{2}\sum\limits_{{\bf
q}\neq0}\varepsilon_q\left[\frac{\alpha_q}{\sh(\beta E_q)}-
 \frac{1}{\sh(\beta \varepsilon_q)}\right]+
 \]\vspace*{-7mm}
\begin{align}\label{Energy}
 +\,\frac{1}{16}\sum\limits_{{\bf q}\neq0}\varepsilon_q\left(\!1-\frac{1}{\alpha_q^2}\!\right)
 \left(\!\alpha_q-\frac{1}{\alpha_q}-4\alpha_q^2\!\right)\!\!,
\end{align}
where $\bar{m}$ is the effective mass; $\bar{\varepsilon}_{q}=\hbar^{2}%
q^{2}/2\bar{m}$, $\bar{z}_{0}$, and $\bar{S}_{0}(q)$ are the renormalized
one-particle spectrum, activity, and structure factor, respectively, of the
ideal Bose gas; $E_{q}=\alpha_{q}\varepsilon_{q}$ is the spectrum of
elementary Bogolyubov excitations; $\alpha_{q}=\sqrt{1+2N\nu_{q}%
/(V{\varepsilon}_{q})}$ is the Bogolyubov factor, $\nu_{q}=\int
e^{-i\mathbf{qR}}\Phi(R)\,d\mathbf{R}$ is the Fourier transform of the pairwise
interparticle interaction potential $\Phi(R)$,
%2
\begin{align}
 S(q)={\bar{S}_0(q)\over 1+\lambda_q\bar{S}_0(q)}
\end{align}
is the structure factor of a Bose liquid in the pair correlation
approximation;
and%
%3
\begin{align}
 \lambda_q=\alpha_q\th\left[\frac{\beta}{2}E_q\right]-\th\left[\frac{\beta}{2}\varepsilon_q\right]\!\!.
\end{align}
The distribution of Bose particles with the new spectrum looks like%
%4
\begin{equation}\label{renorm spectr}
\bar{n}_p={1\over {\bar{z}_0}^{-1}e^{\beta\bar{\varepsilon}_{p}}-1},
\end{equation}
whereas the renormalized one-particle spectrum
$\bar{\varepsilon}_{p}$ is chosen in the form
%5
\begin{eqnarray}\label{renorm spectr energ}
\bar{\varepsilon}_{p}=\varepsilon_{p}+\Delta_{p}-\Delta_{0},
\end{eqnarray}
where $\Delta_{p}$ is a correction to the spectrum, which is to be
determined.\,\,The value of $\Delta_{0}$ depends only on the
temperature and is actually responsible for the activity
renormalization.\,\,After eliminating infra-red divergences, the
expression for $\Delta_{p}$ looks like \cite{VP2008}
%6
\begin{eqnarray}\label{Delta_q}
\Delta_{p}={1\over N\beta}\mathop{\sum_{{\bf q}\neq0}}
{\lambda_{q}\over 1+\lambda_{q}\bar{S}_0(q)}\bar{n}_{|{\bf p}+{\bf
q}|}.
\end{eqnarray}

Expression (\ref{renorm spectr energ}) for the renormalized
one-particle
spectrum can also be written in the form%
%7
\begin{equation}
\bar{\varepsilon}_p={\hbar^2p^2\over2\bar{m}(p)},
\end{equation}
where the quantity $\bar{m}(p)$ is regarded as the total effective
mass of a particle, which depends on the absolute value of wave
vector $\mathbf{p}$.\,\,This form for the spectrum
$\bar{\varepsilon}_{p}$ was proposed in work \cite{TMP08} in order
to exclude infra-red divergences.\,\,It will be recalled that the
effective mass $\bar{m}$ is formed by many-particle correlations,
starting from four-particle ones, and, generally speaking, it
depends on the momentum $p$.\,\,It is clear that we are interested
in the behavior of $\bar{m}(p)$ as $p\rightarrow0$.\,\,As the total
effective mass, we will understand the quantity
$\bar{m}=\bar{m}(0)$.\,\,In this connection, let us consider the
difference $\Delta_{p}-\Delta_{0}$ as $p\rightarrow0$ in more
details.

At small $p$-values, the renormalized spectrum (\ref{renorm spectr energ}) can be
written in the form \cite{VP2008}
%8
\begin{equation}\label{efm}
\bar{\varepsilon}_p={\hbar^2p^2\over2\bar{m}},
\end{equation}
where\vspace*{-2mm}
%9
\[
 {m^*\over\bar{m}}=1+\frac{1}{2\pi^2\rho}\int\limits_0^\infty
\frac{q^2\lambda_q}{1+\lambda_q\bar{S}_0(q)}\,\times
\]\vspace*{-7mm}
\begin{equation}\label{F_T}
\times\,\overline{n}_q(1+\overline{n}_q)
 \left[\frac{2}{3}\beta\varepsilon_q(1+2\overline{n}_q)-1\right]dq.
\end{equation}
In our theory, we use the following expression for the temperature
dependence of the \textquotedblleft seed\textquotedblright\
effective mass $m^{\ast}$, which was obtained in work
\cite{Past2012}:
%10
\begin{align}\label{mass}
&\frac{m}{m^*}=1-\frac{1}{3N}\sum_{{\bf q} \neq
0}\frac{(\alpha_q-1)^2}{\alpha_q(\alpha_q+1)}\,-\nonumber\\
&-\,\frac{2}{3N}\sum_{{\bf q} \neq 0}\Bigg\{\!
\frac{\alpha^2_q+3}{\alpha^2_q-1}\left[n(\beta
\varepsilon_q)-1/(\beta \varepsilon_q)\right]-\nonumber\\
&-\,\frac{3\alpha^2_q+1}{\alpha_q(\alpha^2_q-1)} \left[n(\beta
E_q)-1/(\beta E_q)\right]+\nonumber\\[-2mm]
&+\,2\left[1/(\beta \varepsilon_q)-\beta \varepsilon_q n(\beta
\varepsilon_q)[1+n(\beta \varepsilon_q)]\right]\!\!\Bigg\}\!,
\end{align}
where the notation {$n(x)=1/(e^{x}-1)$} is used.

It is easy to see that the critical-point divergence on the
right-hand side of equality (\ref{F_T}) originates from the
integrand at small $q$-values.\,\,This singularity is logarithmic,
as will be shown later.\,\,Such a divergence is typical of critical
phenomena.\,\,Our task consists in isolating this singularity and
finding a correct expression for the effective mass.\,\,For this
purpose, let us consider the following equality, which follows from
work \cite{VP2008}:
%11
\begin{equation}\label{FTDp0}
\frac{m^*}{\bar{m}}=1+\lim_{p\rightarrow
0}\frac{\Delta_p-\Delta_0}{\varepsilon_p},
\end{equation}
where\vspace*{-2mm}%
%12
\begin{align}\label{Dp0}
\Delta_p-\Delta_0=\frac{1}{N\beta}\!\sum_{{\bf q}\neq
0}\!\frac{\lambda_q}{1+\lambda_q\bar{S}_0(q)}\left\{\bar{n}_{|\bf{q}+\bf{p}|}-\bar{n}_q\right\}\!.
\end{align}
In expression (\ref{Dp0}), we isolate the quantity
%13
\begin{align}
\label{Dinf} &\Delta_{\infty}\!=\!\frac{1}{N\beta}\sum_{{\bf q}\neq
0}\frac{\lambda_q}{1+\lambda_q\bar{S}_0(q)}
\bigg\{\!\frac{1}{\bar{z}_0^{-1}-1+\bar{z}_0^{-1}\beta\bar{\varepsilon}_{|\bf{q}+\bf{p}|}}\,-\nonumber\\
&-\,\frac{1}{\bar{z}_0^{-1}-1+\bar{z}_0^{-1}\beta\bar{\varepsilon}_{q}}\!\bigg\},
\end{align}
which contains the indicated non-analyticity in whole and,
simultaneously, is much more convenient for the analysis.\,\,The
next step consists in finding a series expansion for
$\Delta_{\infty}$ in the interval of small $p$-values and confining
the series to terms proportional to $p^{2}$, because the
higher-order terms give no contribution to the effective mass owing
to equality (\ref{FTDp0}).

On the right-hand side of equality (\ref{Dinf}), we change from summation to
integration:%
%14
\begin{align}\label{Dinfint}
&\Delta_{\infty}=\frac{p_0^2\bar{z}_0}{4\pi^2\beta\rho
}\int\limits_0^{\infty}\frac{\lambda_qdq}
{1+\lambda_q\bar{S}_0(q)}\,\times\nonumber\\
&\times\left\{\!\frac{q}{2p}\ln\left|\frac{P_0^2+(q+p)^2}{P_0^2+(q-p)^2}\right|-\frac{2q^2}{P_0^2+q^2}\!\right\}\!\!,
 \end{align}
where $p_{0}^{2}=2\bar{m}/(\beta\hbar^{2})$ and
$P_{0}=p_{0}\sqrt{(1-\bar {z}_{0})}$.\,\,In the subcritical region,
the activity $\bar{z}_{0}=1$.\,\,Therefore, $P_{0}=0$ here, and the
expressions obtained above become a little simpler.

Let us change the variables: $q/p=x$ and $dq=pdx$. Then%
%15
\begin{align}\label{Dinfint}
&\Delta_{\infty}=\frac{p_0^2\bar{z}_0p}{4\pi^2\beta\rho}\int\limits_{0}^{\infty}
\frac{\lambda_{px}}{1+\lambda_{px}\bar{S}_0(px)}\,\times\nonumber\\
&\times\left\{\!\frac{x}{2}\ln\left|\frac{P_0^2/p^2+(x+1)^2}{P_0^2/p^2+(x-1)^2}
\right|-\frac{2x^2}{P_0^2/p^2+x^2}\!\right\}dx.
\end{align}

\begin{figure}
\vskip1mm
\includegraphics[width=\column]{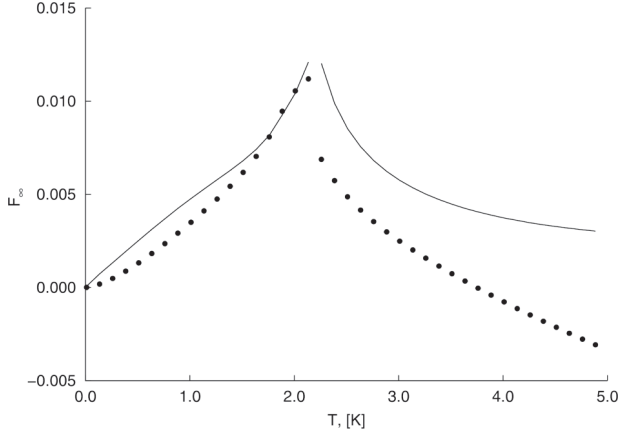}
\vskip-3mm\caption{Temperature dependence of the quantity
{$F_{\infty}=$ =~$\lim
\limits_{p\rightarrow0}\Delta_{\infty}(p)/\varepsilon_{p}$} at
$p=0.01$. Points correspond to the exact expression, and the solid
curve to the accepted approximation }
\end{figure}

\begin{figure}
\vskip3mm
\includegraphics[width=\column]{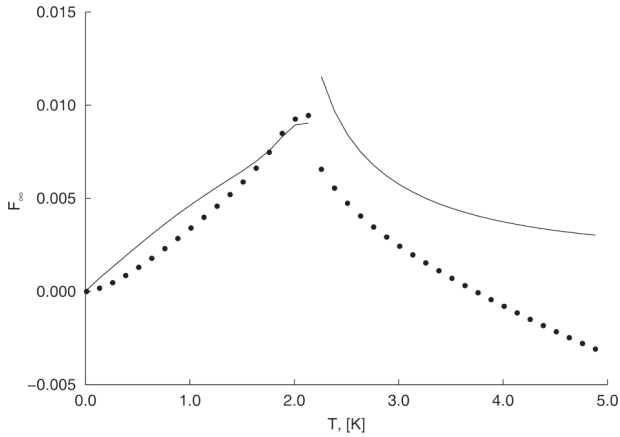}
\vskip-3mm\caption{The same as in Fig.~1, but for $p=0.1$ }
\end{figure}

\noindent
The function $\lambda_{px}/(1+\lambda_{px}\bar{S}_{0}(px))$ is
finite and tends to zero, as $x\rightarrow\infty$ (at a fixed
$p$).\,\,On the other hand,
the function%
\[
\frac{x}{2}\ln\left|\frac{P_0^2/p^2+(x+1)^2}{P_0^2/p^2+(x-1)^2}\right|-\frac{2x^2}{P_0^2/p^2+x^2}\nonumber
\]
falls down to zero in the interval $x>1$.\,\,Moreover,
\[
\int\limits_{0}^{\infty}\left\{\!\frac{x}{2}\ln\left|\frac{P_0^2/p^2+(x+1)^2}{P_0^2/p^2+(x-1)^2}\right|-
\frac{2x^2}{P_0^2/p^2+x^2}\!\right\}dx=0\nonumber
\]
for any $P_{0}$- and $p$-values.\,\,Those facts allow us to assert
(especially if the matter concerns the critical region) that only
the interval, where the $px$-values in the integrand are small,
makes a non-disappearing contribution to the quantity
$\Delta_{\infty}/p^{2}$ as $p\rightarrow0$.

Now, let us proceed to expanding the quantities $\lambda_{q}$ and $\bar{S}%
_{0}(q)$ in a vicinity of the zero wave vector, $q=0$.\,\,For
$\lambda_{q}$,
the result is obtained easily:%
%16
\begin{equation}\label{lr}
\lambda_q=\beta\rho\nu_0+o(q),
\end{equation}
where $\rho$ is the density of the Bose system. On the contrary, the
form of the series expansion for the structure factor of the ideal
Bose gas depends on the
temperature interval.\,\,It looks like%
 %17
\begin{eqnarray}\label{S0r}
S_0(q)=\frac{4\bar{m}(1\!-\!\left(T/T_{\rm
c}\right)^{3/2})}{\beta\hbar^2q^2}\!+\!\frac{{\bar{m}}^2}{2\rho\hbar^4\beta^2}\frac{1}{q}\!+\!1\!
+\!o\left(q\right)
\end{eqnarray}
in the subcritical region ($T<T_{\rm c}$) and
%18
\begin{eqnarray}\label{S0nk}
S_0(q)=\frac{{\bar{m}}^2}{\pi\rho\hbar^4\beta^2}\frac{1}{q}\arctg\left(\!\frac{q}{2P_0}\!\right)+o(q)
 \end{eqnarray}
in the supercritical one.\,\,This result follows immediately from
the expression for the structure factor of the ideal Bose gas
\cite{Vakarchuk04}
\[
\bar{S}_0(q)=1+\frac{\bar{m}}{4\pi^2\rho \beta\hbar^2}\frac{1}{q}
\int\limits_{0}^{\infty}\frac{p}{\bar{z}_0^{-1}e^{\beta\frac{\hbar^2q^2}{2\bar{m}}}-1}\,\times
\]\vspace*{-7mm}
\[
\times\,
\ln\left|\frac{1-\bar{z}_0e^{-\beta\frac{\hbar^2(p+q)^2}{2\bar{m}}}}{1-\bar{z}_0
e^{-\beta\frac{\hbar^2(p-q)^2}{2\bar{m}}}}\right|dp.
\]

In order to analyze expression (\ref{Dinfint}) analytically, let us
apply such approximations for the quantities $\lambda_{q}$ and
$\bar{S}_{0}(q),$ which contain only the expansion terms presented
above.\,\,One can check the adequacy of those approximations with
the help of the numerical analysis, the results of which are
exhibited in Figs.~1 and~2.

The further analysis of expression (\ref{Dinfint}) will be carried out
separately in the sub- and supercritical temperature intervals and at the very
critical point, because the approaches, which should be applied in each of
those cases, are different.

\section{Calculations in the Subcritical Temperature Interval}

In the subcritical temperature interval ($T<T_{\rm c}$), expression (\ref{Dinfint}%
) reads%
\[
\Delta_{\infty}=\frac{p_0^2\varkappa
p^3}{4\pi^2\beta\rho}\!\int\limits_{0}^{\infty}\!
\frac{x^2\left(\!x\ln\left|\frac{x+1}{x-1}\right|-2\!\right)dx}{(1+\varkappa)x^2p^2+\gamma
xp+2n_0\varkappa} +o(p^2),
\]
where%
%19
\begin{equation}
 \varkappa=\beta\rho\nu_0;\quad \gamma=\frac{{\bar{m}}^2\nu_0}{2\hbar^4\beta};\quad
 n_0=1-\left(\!\frac{T}{T_{\rm c}}\!\right)^{\!\!{3}/{2}}\!.
 \end{equation}
Applying the formula
\begin{equation}\label{predst}
x\ln\left|\frac{x+1}{x-1}\right|\!-\!2\!=\!
\int\limits_{-1}^{1}\frac{x^2da}{x^2-a^2}-\int\limits_{-1}^{1}da\!=\!\int\limits_{-1}^{1}
\frac{a^2da}{x^2-a^2}\nonumber
\end{equation}
and changing the order of integration, we obtain%
%20
 \[
\Delta_{\infty}\!=\!\frac{2\bar{m}\varkappa
p^3}{(2\pi\hbar\beta)^2\rho}\int\limits_{-1}^{1}da\,\times
 \]\vspace*{-7mm}
\begin{align}
\label{DKK} \times\int\limits_{0}^{\infty}
\!\frac{x^2a^2dx}{(x^2-a^2)[(1+\varkappa)x^2p^2+\gamma xp
+2n_0\varkappa]}+o(p^2).
\end{align}
The denominator of the integrand should be factorized, and the whole
integrand should be expanded in simple fractions.\,\,Then the
elementary integration over the variable $x$ gives\vspace*{-1mm}
 %21
 \[
\Delta_{\infty}\!\!=\!\!-\frac{p_0^2\varkappa p}{4\pi^2\beta\rho
(1+\varkappa)}\int\limits_{-1}^{1}a^2da
\bigg\{\frac{p^2a\ln|a|}{2(ap-x_1)(ap-x_2)}\,-
 \]\vspace*{-7mm}
 \[
-\,\frac{p^2a\ln|a|}{2(ap+x_1)(ap+x_2)}+\!\frac{px_2^2\ln|x_2/p|}{(a^2p^2-x_2^2)(x_1-x_2)}\,-
 \]\vspace*{-7mm}
\begin{align}\label{intpr}
-\frac{px_1^2\ln|x_1/p|}{(a^2p^2-x_1^2)(x_1-x_2)}\!\bigg\}+o(p^2),
\end{align}
where $x_{1}/p$ and $x_{2}/p$ are roots of the quadratic
equation\vspace*{-1mm}
%21a
\begin{equation}
{(1+\varkappa)x^{2}p^{2}+\gamma xp+2n_{0}\varkappa=0},\label{sqequ}%
\end{equation}
and\vspace*{-3mm}
%22
 \begin{eqnarray}\label{x12}
  x_{1,2}=\frac{-\gamma\pm\sqrt{\gamma^2-8n_0(1+\varkappa)\varkappa}}{2(1+\varkappa)}.
 \end{eqnarray}
After the corresponding transformations and the integration over the
variable $a$, we
obtain%
 %23
\begin{align}\label{Delta_ost}
&\Delta_{\infty}=\frac{p_0^2\varkappa}{4\pi^2\beta\rho
(1+\varkappa)}\frac{1}{x_1-x_2}\bigg\{\!\!\frac{}{}2(x_2^2-x_1^2)\,+\nonumber\\
&+\,\frac{x_2^3}{p^2}\left(\!\dilog\left[1+\frac{p}{x_2}\right]-\dilog\left[1-\frac{p}{x_2}\right]\!\right)-\nonumber\\
 &-\,\frac{x_1^3}{p^2}\left(\!\dilog \left[1+\frac{p}{x_1}\right]-\dilog\left[1-\frac{p}{x_1}\right]\!\right)+\nonumber\\
&+\,2 x_2^2\ln|x_2/p|\left(\!1-\frac{x_2}{p} \arcth\left[\frac{p}{x_2}\right]\!\right)-\nonumber\\
&-\,  2x_1^2\ln|x_1/p|\left(\!1-\frac{x_1}{p}
\arcth\left[\frac{p}{x_1}\right]\!\right)\!\!\bigg\}+o(p^2),
\end{align}
where%
\[
\mathrm{dilog}[x]=\int\limits_{1}^{x}\ln(y)/(1-y)dy.
\]
Expanding the obtained expression in $p$, we find that the quantities
proportional to $p^{2}$ originate exclusively from the last two terms in
the braces.\,\,As a result, we obtain%
%24
\begin{align}
\Delta_{\infty}=\frac{p_0^2\varkappa}{3\pi^2\beta\rho
(1+\varkappa)}\frac{\ln\left|x_1/x_2\right|}{(x_1-x_2)}p^2+o(p^2).
\end{align}

When approaching the critical point, one of the roots, say $x_{2}$,
tends to zero, and we obtain a logarithmic divergence for the
quantity $\Delta_{\infty }$ in a vicinity of the critical
point.\,\,What is the effective mass in this case? Returning to the
analysis of the expressions for $x_{1}$ and $x_{2}$, we may conclude
that there exists a temperature $T_{F},$ at which $x_{1}$ and
$x_{2}$ are real-valued quantities.\,\,In this case, the function
$\mathrm{arc\tanh}\left(  p/x_{2}\right)$ is no more finite in the
temperature interval between $T_{F}$ and $T_{\rm c}$ and diverges
when approaching the critical point.\,\,Until the quantity $x_{2}$
remains complex, the $\mathrm{arc\tanh}$ function can be expressed
in terms of the trigonometric arctangent, which is finite.\,\,The
temperature $T_{F}$ can be easily found by putting the discriminant
of the quadratic equation (\ref{sqequ}) equal to zero.\,\,Its
numerical solution gives $T_{F}\approx2.13$~K if the critical
temperature $T_{\rm c}\approx2.18$~K.\,\,One can see that this is a
very narrow interval, which can be interpreted as a fluctuation one,
i.e. when the fluctuations of the Bose condensate becomes comparable
with its amount.\,\,It can also be considered as a region similar to
the Ginzburg region, where the perturbation calculation method
fails.\,\,In any case, in the framework of our approach, we cannot
draw any proper conclusion about the effective mass in this narrow
interval.\,\,Other methods, e.g., the renorm-group approach, are
required to analyze this region.\,\,The numerical analysis testifies
that the contribution $\Delta_{\infty}$ to the effective mass is
very insignificant at temperatures below $T_{F}$.\,\,The analytical
form of this contribution to the right-hand side of
Eq.~(\ref{FTDp0}) is as follows:%
%25
 \begin{align}
 \frac{p_0^4\varkappa}{6\pi^2\rho
(1+\varkappa)}\frac{\ln\left|x_1/x_2\right|}{(x_1-x_2)}.
 \end{align}

\section{Calculations at Critical Point}

In order to elucidate the divergence character of the quantity
$\Delta _{\infty}/p^{2}$, regarded as a function of $p$, at the
critical point, let us make calculations in this case.\,\,Let us
return to formula (\ref{DKK}) and put
$T=T_{\rm c}$ in it, which means that $n_{0}=0$.\,\,Then,%
 %26
\begin{align}
\Delta_{\infty}\!=\!\frac{p_0^2\varkappa
p^2}{2\pi^2\beta\rho}\int\limits_{0}^{1}\!\!da\!\!
\int\limits_{0}^{\infty}
\!\!\frac{xa^2dx}{(x^2-a^2)[(1+\varkappa)xp+\gamma]}+o(p^2).
\end{align}

Again, let us factorize the denominator of the integrand, expand the
resulting integrand in simple fractions, and integrate over the
variable $x$.\,\,As a result, we obtain
\begin{eqnarray}
\Delta_{\infty}
 =\frac{p_0^2\varkappa p^2}{2\pi^2\beta\rho(1+\varkappa)}\int\limits_0^1
 \frac{a^2x_0\ln|a/x_0|}{a^2p^2-x_0^2}da+o(p^2),\nonumber
\end{eqnarray}
where $x_{0}/p$ is a root of the equation
\[
{(1+\varkappa)xp+\gamma=0},%
\]
and $x_{0}=-\gamma/(1+\varkappa)$.\,\,Changing the variables,
{$a/|x_{0}|=\xi$},
in the integral above, we have%
%27
\[
\Delta_{\infty}=\frac{p_0^2\varkappa
x_0^2p}{2\pi^2\beta\rho(1+\varkappa)}\int\limits_0^{1/|x_0|}
 \frac{\xi^2\ln\xi}{\xi^2-1}+o(p^2)=
\]\vspace*{-6mm}
\begin{align}
 =-\frac{p_0^2\varkappa x_0^2p}{2\pi^2\beta\rho(1+\varkappa)}\int\limits_0^{1/|x_0|}
 \xi^2\ln\xi d\xi+o(p^2),
 \end{align}
since $1/\left\vert x_{0}\right\vert \sim p$ ($p\rightarrow0$). As a result,
we obtain
\begin{eqnarray}
 \Delta_{\infty}\!=\frac{p_0^2\varkappa p^2}{18\pi^2\beta\rho(1+\varkappa)|x_0|}
 \left(\!1-3\ln\left|\frac{p}{x_0}\right|\!\right)\!+\!o(p^2).\nonumber
\end{eqnarray}

Hence, we showed that the quantity $\Delta_{\infty}/p^{2}$ diverges
at the critical point as $\ln\left\vert p\right\vert $
($p\rightarrow0$).\,\,Such a singularity is typical of critical
phenomena.\,\,It can be interpreted as a consequence of the
expansion of the one-particle spectrum of a Bose liquid in
a vicinity of the critical point:%
%28
\begin{align}
&\frac{\!\hbar^2
\tilde{p}^2}{2\bar{m}}\left(\frac{p}{\tilde{p}}\!\right)^{\!\!2-\eta}=
\frac{\hbar^2\tilde{p}^2}{2\bar{m}}\left(\!\frac{p}{\tilde{p}}\!\right)^{\!\!2}e^{-\eta\ln\left(p/\tilde{p}\right)}=\nonumber\\
&=\frac{\hbar^2
p^2}{2\bar{m}}\left(\!1-\eta\ln\left(\!\frac{p}{\tilde{p}}\!\right)\!\right)+o\left(\eta\right),
 \end{align}
where $\eta$ is the small critical index, and $\tilde{p}$ a
characteristic scale of the wave vector in a vicinity of the
critical point.\,\,Taking into account that only the quantity
$\Delta_{\infty}$ gives a non-zero contribution to the one-particle
spectrum of a Bose liquid at the critical point, we obtain the
following equation for the determination of $\eta$ and $\tilde{p}$:
%29
\begin{align}
&\frac{p_0^2\varkappa p^2}{18\pi^2\beta\rho(1+\varkappa)|x_0|}
\left(\!1-3\ln\left|\frac{p}{x_0}\right|\!\right)=\nonumber\\
&=\frac{p^2
}{p_0^2\beta}\left(\!1-\eta\ln\left(\!\frac{p}{\tilde{p}}\!\right)\!\right)\!\!.
\end{align}
From whence, we have%
%30
\begin{equation}
\begin{array}{l}
\displaystyle  \eta=\frac{4}{3\pi^2}\approx0.135,\\[3mm]
\displaystyle
\tilde{p}=|x_0|\exp\left(\!\frac{\eta-3}{3\eta}\!\right)\approx1.68\cdotp10^{-3}~\text{\AA}^{-1}.
\end{array}
\end{equation}

The result for the small critical index $\eta$ was obtained for the
first time in works \cite{Ferrell, Abe}.\,\,The cited authors used a
method of expansion in reciprocal powers of the order parameter
dimensionality.\,\,The random-phase approximation reproduces only
the first term of this expansion.\,\,Therefore, it is no wonder that
the result obtained for the small critical index differs from the
result of Monte-Carlo simulations \cite{Campostrini}.

\section{Calculations at Above-Critical Temperatures}

At temperatures higher than the critical one, quantity (\ref{Dinfint})
acquires the form%
%31
\[
\Delta_{\infty}=\frac{p_0^2\bar{z}_0\varkappa}{4\pi^2\beta\rho}\int\limits_0^{\infty}
 \frac{qdq}{q+\tilde{\gamma}\arctg\left(\!\frac{q}{2P_0}\!\right)}\,\times
\]\vspace*{-7mm}
\begin{align}
\times\left\{\!\frac{q}{2p}\ln\left|\frac{P_0^2+(q+p)^2}{P_0^2+(q-p)^2}\right|-\frac{2q^2}{P_0^2+q^2}\!\right\}\!\!,
 \end{align}
where $\tilde{\gamma}=2\gamma/\pi$.\,\,Differentiating it with
respect to $p$, integrating the result again over $p$, and changing
the order of integration, we obtain
%32
\begin{align}
  &\Delta_{\infty}=\frac{p_0^2\bar{z}_0\varkappa}{4\pi^2\beta\rho}\frac{1}{p}\int\limits_0^p dp\int\limits_0^{\infty}
 \frac{qdq}{q+\tilde{\gamma}\arctg\left(\!\frac{q}{2P_0}\!\right)}\,\times\nonumber\\
&\times\frac{2qp^2(q^2-p^2-3P_0^2)}{\left[P_0^2+(p+q)^2\right]\left[P_0^2+(p-q)^2\right]\left[P_0^2+q^2\right]}.
\end{align}
In order to calculate this integral, we symmetrize the limits of
integration over $q$, make an analytical continuation of the
integrand into the upper half-plane of the complex $q$-variable, and
close the contour of integration by a semicircle of radius
$R$.\,\,In the limit $R\rightarrow\infty$, the integral along the
semicircle $R$ equals zero, because the power of the integrand's
denominator is larger by two than the power of the numerator.\,\,As
a result, our integral is equal to a sum of residues at the
analytical continuation of the integrand into the upper half-plane
times $2\pi i$.\,\,Only three singular points of the integrand fall
within this half-plane: $q=p+iP_{0}$, $q=-p+iP_{0}$, and
$q=iP_{0}$.\,\,(Note, by the way, that the multiplier in the
denominator with the $\arctan$ function does not equal to zero over
the whole complex $q$-plane.) As a result of calculations, we obtain
%33
\begin{align}
&\Delta_{\infty}=\frac{p_0^2\bar{z}_0\varkappa}{4\pi^2\beta\rho}\frac{\pi
i}{2p}\int\limits_0^p dp \Bigg(\!
\frac{2P_0^2}{iP_0+\tilde{\gamma}\arctg(i/2)}\,+\nonumber\\
&+\frac{(-p+i P_0)^2}{-p+i
P_0+\tilde{\gamma}\arctg\left(\!-\frac{p}{2P_0}
+\frac{i}{2}\!\right)}\,+\nonumber\\
& +\frac{(p+i P_0)^2}{p+i
P_0+\tilde{\gamma}\arctg\left(\!\frac{p}{2P_0}
+\frac{i}{2}\!\right)} \!\Bigg)\!.
\end{align}

Without specifying the subsequent rather simple transformations, we present the
final result for $\Delta_{\infty}$:%
%34
\begin{align}
&\Delta_{\infty}\!\!=\!\!-\frac{2\pi}{p}\!\int\limits_0^p\! dp
\left\{\!\frac{(f_2\!+\!P_0)(p^2\!-\!P_0^2)\!-\!2pP_0(f_1+p)}{(f_1+p)^2+(f_2+P_0)^2}\,+\right.\nonumber\\
&\left. +\,\frac{P_0^2}{P_0+\tilde{\gamma}\ln(3)/2}\!\right\}\!\!,
\end{align}
where%
%35
\begin{equation}
\begin{array}{l}
\displaystyle
f_1=\frac{\tilde{\gamma}}{2}\arctg\left(\!\frac{4pP_0}{3P_0^2-p^2}\!\!\right)\!\!,\\[3mm]
\displaystyle
f_2=-\frac{\tilde{\gamma}}{2}\ln\left(\!\!\frac{\sqrt{(3P_0^2-p^2)^2}+16p^2P_0^2}{9P_0^2+p^2}\!\!\right)\!\!.
\end{array}
\end{equation}
We expand the expression obtained in a series in the small parameter $p$
and keep
only the terms proportional to $p^{2}$. As a result, we obtain%
%36
\begin{align}
 &\Delta_{\infty}=-\frac{p_0^2\bar{z}_0\varkappa\tilde{\gamma}}{27\pi\beta\rho}\,\times\nonumber\\
 &\times\frac{\left(-9\tilde{\gamma}\ln^2(3)+8P_0+28\tilde{\gamma}\ln(3)-16\tilde{\gamma}\right)}
 {(2P_0+\tilde{\gamma}\ln(3))^3}p^2+o(p^2).
\end{align}%\vspace*{-3mm}

\noindent The corresponding contribution to the right-hand side of
Eq.~(\ref{FTDp0}) is
as follows:%
%37
\begin{align}
 -\frac{p_0^4\bar{z}_0\varkappa\tilde{\gamma}}{27\pi\rho}
\frac{\left(-9\tilde{\gamma}\ln^2(3)+8P_0+28\tilde{\gamma}\ln(3)-16\tilde{\gamma}\right)}
 {(2P_0+\tilde{\gamma}\ln(3))^3}.\!\!\!\!\!
\end{align}%\vspace*{-5mm}

\noindent With the help of the numerical analysis, one can get
convinced in the smallness of this quantity.\,\,Therefore, its
contribution to the effective mass can also be neglected.

\section{Analytical Expression for Effective Mass}

Taking into account that the quantity $\Delta_{\infty}$ gives an insignificant
contribution to the effective mass, which was demonstrated above, and
returning to the calculation scheme described in work \cite{VP2008}, we obtain
the following expression for the effective mass:%
%38
\begin{equation}
\bar{m}=\frac{m^*}{\left(1+F(T)\!\right)},
\end{equation}\vspace*{-5mm}
where%
%39
\begin{align}
 &F(T)=\lim\limits_{p\rightarrow 0}
 \frac{1}{N\beta\varepsilon_p}\sum_{{\bf q}\neq 0}\frac{\lambda_q}{1+\lambda_q\bar{S}_0(q)}
 \left(e^{{\bf p}{\bf \nabla}_q}\!-\!1\right)\times\nonumber\\[-1.5mm]
 &\times\left(\!\bar{n}_q-\frac{1}{\bar{z}_0^{-1}(\beta\bar{\varepsilon}_q+1-\bar{z}_0)}\!\right)\!\!,
 \end{align}\vspace*{-5mm}

\noindent and $\mathbf{\nabla}_{q}$ is the gradient operator.

Let us expand the operator
$e^{\mathbf{p}\mathbf{\bigtriangledown}_{q}}$ in a series and
confine the expansion to first three terms, because they give us the
required approximation.\,\,Making simple transformations, changing
from summation to integration, and taking the meaning of notations
$p_{0}$ and $P_{0}$ into account, we obtain the following expression
for the quantity
$F(T)$:%
%40
\begin{align}
&F(T)=\frac{1}{2\pi^2\rho}\int\limits_0^{\infty}\frac{\lambda_qq^2dq}{1+\lambda_q\bar{S}_0(q)}
\Bigg(\!\!\frac{}{}\overline{n}_q(1+\overline{n}_q)\,\times\nonumber\\
&\times\left[\frac{2}{3}\beta\varepsilon_q(1+2\overline{n}_q)-1\right]
-\frac{\bar{z}_0(\beta\bar{\varepsilon}_q-3+3\bar{z}_0)}{3
\left(\beta\bar{\varepsilon}_q+1-\bar{z}_0\right)^3}\!\!\Bigg)\!.
\end{align}
A direct inspection easily verifies that the function $F(T)$ equals
zero in the limits of both low and high temperatures.\,\,Therefore,
in those limits, $\bar{m}=$ $=m^{\ast}$.\,\,Using the results of
work \cite{Past2012}, we obtain that
$\lim\limits_{T\rightarrow0}\bar{m}\approx1.7m$ and $\lim
\limits_{T\rightarrow\infty}\bar{m}=m$.

\section{Numerical Calculation of~Effective~Mass~and~Heat~Capacity}

\begin{figure}[b!]
\vskip-2mm
\includegraphics[width=\column]{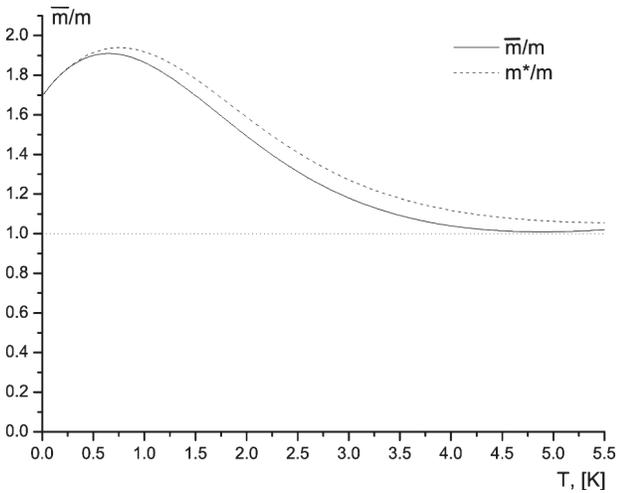}
\vskip-3mm\caption{ Temperature dependence of the effective mass of
$^{4}$He atom }
\end{figure}

\begin{figure}[b!]
 \vskip2mm
\includegraphics[width=\column]{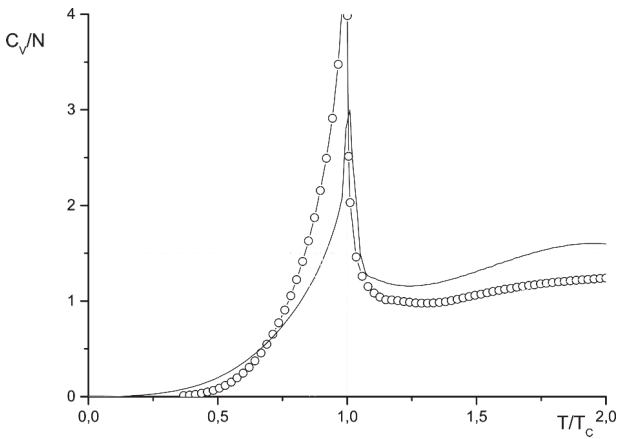}
\vskip-3mm\caption{Temperature dependence of the heat capacity of
liquid $^{4}$He: (solid curve)~theoretical result obtained taking
the effective mass into account, (circles)~experimental data
{\cite{Ceperley,Arp1,Arp2}} }\vspace*{1.5mm}
\end{figure}

Let us illustrate the obtained result in the graphic form.\,\,For
this purpose, we should numerically calculate the ratio
$\overline{m}/m$.\,\,The corresponding calculation is
self-consistent, because the expression for $\overline{m}$ includes
the quantities $\bar{S}_{0}(q)$, $\bar{\varepsilon}_{q}$, and
$\overline{n}_{q}$, which depend, in turn, on $\overline{m}$.\,\,In
practice, this situation implies the application of an iteration
process, which took 3--4 cycles in our case.

The calculations were carried out for the equilibrium helium density
$\rho=0.02185$~{\AA}$^{-3}$, the particle mass $m=4.0026$~amu, and
the sound velocity $c=238.2$~m/s in the limit $T\rightarrow0$
\cite{Donnelly98}.\,\,The experimentally measured structure factor
$S^{\mathrm{exp}}(q)$ for liquid $^{4}$He extrapolated to the
temperature $T\rightarrow0$ \cite{Rovenchak00} rather than the
Fourier coefficient for the energy of pairwise interparticle
interaction $\nu_{q}$ was used as the input information.

In Fig.\,\,3, the temperature dependence of the effective mass of
$^{4}$He atom calculated in the approximation of pair interparticle
correlations is exhibited.\,\,On its base, using the known formula
\cite{Vstup}, we also calculated the temperature of the Bose
condensation in liquid $^{4}$He.\,\,The obtained value is $T_{\rm
c}\approx2.18$~K, which is very close to the experimental value
$T_{\rm c}=2.168$~K.

While calculating the heat capacity, we used expression
(\ref{Energy}) for the internal energy of a many-boson system in the
pair correlation approximation.\,\,We numerically differentiated it
with respect to the temperature.\,\,Figure~4 demonstrates the
temperature dependence of the heat capacity calculated with regard
for the effective mass.

\section{Conclusions}

An expression for the temperature dependence of the effective mass
of a $^{4}$He atom (in both the normal and superfluid phases) is
obtained.\,\,It allows infra-red divergences, which are typical of
critical phenomena, to be eliminated.\,\,The expression for the
effective mass is applicable at all temperatures, except for a
narrow fluctuation interval between the temperature
$T_{F}\approx2.13$ K and the temperature of phase transition.\,\,In
the high-temperature limit, as well as when the interparticle
interaction is \textquotedblleft switched-off\textquotedblright, the
effective mass transforms into the \textquotedblleft
seed\textquotedblright\ mass of a $^{4}$He atom.\,\,In the
low-temperature limit, we obtain a value that coincides with the
effective mass of a $^{3}$He impurity atom in liquid $^{4}$He,
provided that the \textquotedblleft seed\textquotedblright\ mass of
a $^{3}$He atom is substituted by the mass of a $^{4}$He one
\cite{Vakar96}.\,\,In this context, we note that there is no common
opinion concerning the effective mass even at the zero temperature,
to say nothing of a wide temperature interval, because the
introduction of this quantity into consideration is a
phenomenological issue and, to a great extent, depends on the
approaches applied for its calculation
\cite{Samulski,visnyk93,Rovenchak03}.

The behavior of the heat capacity curve theoretically calculated
with regard for the effective mass is in much better agreement with
the experimental data than if without it, in particular, in the
supercritical region \cite{VP2007}.\,\,In addition, in comparison
with the \textquotedblleft bare\textquotedblright\ mass, the
effective mass obtained in this work gives a better agreement with
the experimental data for the heat capacity in the temperature
interval of about 0.5$~\mathrm{K}$ above the phase transition point
\cite{Past2012}.

The application of the effective mass made it possible to shift the
phase transition point from the value for the ideal Bose gas to the
temperature $T_{\rm c}\approx2.18$~K.\,\,As was already mentioned,
the latter value is very close to the experimental one.\,\,The
\textquotedblleft bare\textquotedblright\ mass gives rise to $T_{\rm
c}\approx1.94$~K in this case \cite{Past2012}.

In the framework of the approach proposed in this work, we also
succeeded in finding the small critical index $\eta$ in the
random-phase approximation.\,\,The obtained value differs rather
strongly from the recommended one \cite{Campostrini}, but
simultaneously reproduces the well-known result of this
approximation \cite{Ferrell}.

\vspace*{-3mm}
\rezume{%
І.О.\,Вакарчук, О.І.\,Григорчак,\\
В.С.\,Пастухов,Р.О.\,Притула}{ЕФЕКТИВНА МАСА АТОМА $^4$He\\ В
НАДПЛИННІЙ І НОРМАЛЬНІЙ ФАЗАХ} {Знайдено вираз для температурної
залежності
 ефективної маси атома $^4$He в надплинній і нормальній фазах, який дозволяє
 усунути інфрачервоні розбіжності і є застосовним при всіх температурах за винятком
 вузької флуктуаційної області $0{,}97\lesssim T/T_c\leq 1$.
 В границі високих і низьких температур, а також в границі виключення взаємодії, отриманий вираз
 дає відомі результати. На основі ефективної маси розраховано хід кривої теплоємності, а також
 знайдено температуру фазового переходу $T_c\approx2{,}18$ K.
 Використовуючи запропонований в роботі підхід, отримано значення малого критичного індексу $\eta$ в
  наближенні хаотичних фаз, яке відтворює вже відомий результат цього наближення.}


\begin{thebibliography}{99}                                                                                               %

%1
\bibitem {London}F. London, Phys. Rev. \textbf{54}, 947 (1938).
%2
\bibitem {Pashitskii}E.A. Pashitskii, Fiz. Nizk. Temp. \textbf{25}, 115 (1999).
%3
\bibitem {DalfovoGiorginiPitaevskiiStringari}F. Dalfovo, S. Giorgini, L.P.
Pitaevskii, and S. Stringari, Rev. Mod. Phys. \textbf{71}, 463 (1999).
%4
\bibitem {BlochDalibardZwerger}I. Bloch, J. Dalibard, and W. Zwerger, Rev.
Mod. Phys. \textbf{80}, 885 (2008).
%5
\bibitem {Feynman53}R.P. Feynman, Phys. Rev. \textbf{91}, 1291 (1953).
%6
\bibitem {PieriStrinatiTifrea}P. Pieri, G.C. Strinati, and I. Tifrea, Euro.
Phys. J. B \textbf{22}, 79 (2001).
%7
\bibitem {Stoof}H.T.C. Stoof, Phys. Rev. A \textbf{45}, 8398 (1992).
%8
\bibitem {BaymBlaizotHolzmannLaloeVautherin}G. Baym, J.-P. Blaizot, M.
Holzmann, F. Lalo{\"{e}}, and D.~Vau\-therin, Phys. Rev. Lett.
\textbf{83}, 1703 (1999).
%9
\bibitem {HolzmannKrauth}M. Holzmann and W. Krauth, Phys. Rev. Lett.
\textbf{83}, 2687 (1999).
%10
\bibitem {ArnoldMooreTomasik}P. Arnold, G. Moore, and B. Toma\v{s}ik, Phys.
Rev. A \textbf{65}, 013606 (2002).
%11
\bibitem {Kastening}B. Kastening, Phys. Rev. A \textbf{69}, 043613 (2004).
%12
\bibitem {KashurnikovProkofevSvistunov}V.A. Kashurnikov, N.V. Prokof'ev, and
B.V. Svistunov, Phys. Rev. Lett. \textbf{87}, 120402 (2001).
%13
\bibitem {ArnoldMoore}P. Arnold and G. Moore, Phys. Rev. Lett. \textbf{87},
120401 (2001).
%14
\bibitem {ReppyCrookerHebralCorwinHeZassenhaus}J.D. Reppy, B.C. Crooker, B.
Hebral, A.D. Corwin, J. He, and G.M. Zassenhaus, Phys. Rev. Lett. \textbf{84},
2060 (2000).
%15
\bibitem {Samulski}A. Isihara and T. Samulski, Phys. Rev. B \textbf{16}, 1969 (1977).
%16
\bibitem {visnyk93}I.O.~Vakarchuk, Visn. Lviv. Univ. Ser. Fiz. \textbf{26}, 29 (1993).
%17
\bibitem {Rovenchak03}A.A. Rovenchak, Fiz. Nizk. Temp. \textbf{29}, 145 (2003).
%18
\bibitem {Vakar96}I.O.~Vakarchuk, J. Phys. Stud.  \textbf{1}, 25 (1996).
%19
\bibitem {Vakar97}I.O. Vakarchuk, J. Phys. Stud. \textbf{1}, 156 (1997).
%20
\bibitem {Lindenau99}M.L. Ristig, T. Lindenau, M. Serhan, and J.W. Clark, J.~Low Temp. Phys. \textbf{114}, 317 (1999).
%21
\bibitem {Lindenau02}T. Lindenau, M.L. Ristig, J.W. Clark, and K.A. Gernoth,
J. Low Temp. Phys. \textbf{129}, 143 (2002).
%22
\bibitem {Gernoth08}K.A. Gernoth, M. Serhan, and M.L. Ristig, Phys. Rev. B
\textbf{78}, 054513 (2008).
%23
\bibitem {Gernoth09}K.A. Gernoth and M.L. Ristig, Int. J. Mod. Phys. B
\textbf{23}, 4096 (2009).
%24
\bibitem {VP2008}I.O.~Vakarchuk and R.O.~Prytula, J. Phys. Stud.  \textbf{12},
4001 (2008).
%25
\bibitem {Past2012}I.O. Vakarchuk, V.S. Pastukhov, and R.O. Prytula, Ukr. J.
Phys. \textbf{57}, 1214 (2012).
%26
\bibitem {Vakarchuk04}I.O. Vakarchuk, J. Phys. Stud. \textbf{8}, 223 (2004).
%27
\bibitem {TMP08}I.O. Vakarchuk, Theor. Math. Phys. \textbf{154}, 6 (2008).
%28
\bibitem {Ferrell}R.A. Ferrell and D.J. Scalapino, Phys. Lett. A \textbf{41},
371 (1972); Phys. Rev. Lett. \textbf{29}, 413 (1972).
%29
\bibitem {Abe}R. Abe, Prog. Theor. Phys. \textbf{49}, 1877 (1973).
%30
\bibitem {Campostrini}M. Campostrini, M. Hasenbusch, A. Pelissetto, and E.
Vicari, Phys. Rev. B \textbf{74}, 144506 (2006).
%31
\bibitem {VP2007}I.O.~Vakarchuk and R.O.~Prytula, J. Phys. Stud.  \textbf{11},
259 (2007).
%32
\bibitem {Donnelly98}R.J. Donnelly and C. F. Barenghi, J. Phys. Chem. Ref.
Data \textbf{27}, 1217 (1998).
%33
\bibitem {Rovenchak00}I.O. Vakarchuk, V.V. Babin, and A.A. Rovenchak, J. Phys.
Stud. \textbf{4}, 16 (2000).
%34
\bibitem {Vstup}I.O.~Vakarchuk, \textit{Introduction to the Many-Body Problem}
(Ivan Franko Lviv National University, Lviv, 1999) (in Ukrainian).
%35
\bibitem {Ceperley}D.M. Ceperley, Rev. Mod. Phys. \textbf{67}, 279 (1995).
%36
\bibitem {Arp1}V.D. Arp, R.D. McCarty, and D. G. Friend, Natl. Inst. Stand.
Technol. Tech. Note 1334 (revised) (1998).
%37
\bibitem {Arp2}V.D. Arp, Int. J. Thermophys. \textbf{26}, 1477 (2005).\vspace*{1mm}
\begin{flushright}
{\footnotesize Received 10.06.15.\\ Translated from Ukrainian by
O.I.~Voitenko}
\end{flushright}
\end{thebibliography}
\end{document}